\documentclass[conference,10pt]{IEEEtran}

\newif\ifarxivversion
\arxivversiontrue

\usepackage{amsmath,amssymb}
\usepackage{booktabs}
\usepackage{graphicx}
\usepackage{multirow}
\usepackage{microtype}
\usepackage{tikz}
\usetikzlibrary{arrows.meta,positioning,shapes.geometric,fit}
\usepackage{xcolor}
\usepackage{url}
\ifarxivversion
  \usepackage[hidelinks,pdfauthor={Bodong Hou and Zhixin Pan},pdftitle={When Is RTL Better? Auditing Evaluation Stability with Tool-Grounded Requirement Profiles}]{hyperref}
\else
  \usepackage[hidelinks,pdfauthor={},pdftitle={When Is RTL Better? Auditing Evaluation Stability with Tool-Grounded Requirement Profiles}]{hyperref}
\fi

\definecolor{profileblue}{RGB}{44,92,143}
\definecolor{profilegreen}{RGB}{48,135,98}
\definecolor{profileorange}{RGB}{213,119,38}
\definecolor{profilegray}{RGB}{104,112,120}
\newcommand{\sys}{\textsc{ReqProfile}}
\newcommand{\uir}{\mathrm{UIR}}
\newif\ifdacreferencespage
\ifarxivversion
  \dacreferencespagefalse
\else
  \dacreferencespagetrue
\fi

\title{When Is RTL Better?\\
Auditing Evaluation Stability with Tool-Grounded Requirement Profiles}

\ifarxivversion
\author{
\IEEEauthorblockN{Bodong Hou and Zhixin Pan\textsuperscript{*}}
\IEEEauthorblockA{Department of Electrical and Computer Engineering,\\
FAMU-FSU College of Engineering, Florida State University,\\
Tallahassee, FL, USA\\
bh24g@fsu.edu; zpan@eng.famu.fsu.edu\\
\textsuperscript{*}Corresponding author: Zhixin Pan}
}
\else
\author{}
\fi

\begin{document}
\maketitle

\begin{abstract}
RTL-generation studies increasingly report functional correctness and
post-synthesis quality, yet the evaluator itself is rarely treated as an
experimental variable. A candidate can appear better when a security rule,
protocol obligation, target context, feasibility gate, uncertainty policy, or
provenance constraint is omitted. We introduce \emph{tool-grounded requirement
profiles}: machine-readable contracts binding each requirement's meaning to
its executable evidence and comparison rule, together with cross-requirement
gates, uncertainty handling, and provenance. We audit evaluator stability by
freezing candidate pairs and perturbing only their evaluation profile. In a
preregistered controlled pilot with six designs, twelve pairs, three
requirement families, six evaluator conditions, and 72 decisions, a generic
cell-count evaluator reports unsupported improvement for 6/12 pairs
(50\%; Wilson 95\% interval 25.4--74.6\%; design-cluster bootstrap interval
25--75\%). The matched evaluator detects all twelve injected regressions,
whereas syntax, generic synthesis, context mismatch, and gate ablation each
lose the required distinction for 12/12 pairs. In a separately frozen external
validation of all 290 public RTLLM v1.1 text-only candidates, 81/215 evaluable
comparisons (37.7\%; task-cluster 95\% interval 24.8--51.4\%) yield unsupported
scalar improvements, including 72 functional-gate reversals. Separately, 26
OpenROAD-flow-scripts runs show why timing context and activity qualification
must be part of the contract. The study does not rank generators; it establishes
evaluator configuration as a measurable, auditable source of conclusion
instability.
\end{abstract}


\section{Introduction}

Compilation is necessary but insufficient evidence that generated RTL is
usable. Recent benchmarks therefore add simulation, formal or reference-based
checks, synthesis, and physical metrics
\cite{lu2024rtllm,pinckney2025cvdp,susan2026grade,fu2026synthesis}. This
progress creates a second problem: an evaluation result is meaningful only
under the requirements and contexts that its evidence actually supports.
Generic cell count cannot discharge a ready/valid obligation; a passing
simulation under one arbitration policy does not prove another; an area value
is not comparable across mismatched target constraints; and a power number is
not decision evidence if its workload activity is invalid.

Most benchmark reporting fixes candidate generation and varies models or
prompts. We instead fix the candidate pair and vary the evaluator. This turns a
normally implicit choice into a controlled factor and asks: \emph{how often
does a plausible but incomplete evaluator support a conclusion that a matched
requirement profile rejects?} We call this \emph{evaluation-induced conclusion
instability}.

Our abstraction, \sys{}, binds requirements, tools, decisions, uncertainty,
and provenance. We freeze a prospective matrix before execution and evaluate
two controlled pairs from each of six paper-only RTL designs. The primary
contrast compares a scalar generic-synthesis evaluator with the matched
profile. Five additional conditions isolate syntax-only, generic, context
mismatch, and distinguishing-gate ablation effects. This is an evaluator audit,
not a leaderboard: the controlled variants provide known regressions, so a
decision can be labeled supported or unsupported without assuming that one
LLM represents a population of generators.

This paper makes four contributions:

\begin{enumerate}
  \item a machine-readable requirement-profile abstraction joining semantics,
  executable evidence, comparison relations, uncertainty, and provenance;
  \item a prospective evaluator-audit protocol that freezes pairs and perturbs
  the evaluation configuration rather than the candidates;
  \item decision-level metrics, including Unsupported Improvement Rate (UIR),
  loss of required distinction, transition counts, and explicit invalid or
  inconclusive outcomes; and
  \item a fully reconstructable 72-decision controlled pilot, a frozen
  290-candidate public text-only external validation, and two non-confirmatory
  physical-design studies illustrating context and evidence safeguards.
\end{enumerate}

We make no claim that six designs estimate the prevalence of evaluator error
in the broader RTL ecosystem, nor that context-dependent PPA is novel. The
claim is narrower: evaluator sufficiency can be specified, perturbed, measured,
and audited before using its output to compare generators.

\subsection{Research Questions and Falsifiable Claims}

We ask four questions. \textbf{RQ1:} How often does a generic scalar evaluator
prefer an injected-regression variant that a matched evaluator rejects?
\textbf{RQ2:} Which conclusion transitions arise when syntax, generic
synthesis, context substitution, or gate ablation replaces the matched
profile? \textbf{RQ3:} Does the same missing-gate mechanism recur in a complete
public cohort of natural text-only LLM outputs? \textbf{RQ4:} Can the same
abstraction represent physical-design
feasibility, repeatability, and evidence-qualification boundaries without
silently forcing a rank?

The preregistered primary estimand for RQ1 is $\uir_{C2}$. The pilot supports
the methodological claim only if at least two requirement families contain a
scalar-to-matched reversal, every reported decision is reconstructable, and
the clean rerun reproduces the analysis. RQ2 retains the full transition matrix
rather than reducing all errors to UIR. RQ3 requires a separately frozen
complete cohort, explicit reference-invalid outcomes, task-clustered
uncertainty, and an exact rerun. RQ4 is non-confirmatory and requires
context-indexed physical records, multiple seeds, and an activity check capable
of invalidating a numeric power result. Thus each claim can fail.

\section{Related Work and Gap}

RTLLM introduced progressive syntax, functionality, and design-quality goals
for natural-language-to-RTL benchmarking \cite{lu2024rtllm}. TuRTLe unifies
syntax, functional, synthesis, PPA, and completion evaluation across existing
benchmarks \cite{garcia2025turtle}; CVDP and RTL-BenchLS broaden task diversity
and design scale \cite{pinckney2025cvdp,wang2026rtlbenchls}. GRADE-RTL separates
port, compilation, elaboration, module-completeness, functional-equivalence,
and implementation checks \cite{susan2026grade}. Synthesis-in-the-loop studies
quantify quality and failure modes using post-synthesis metrics
\cite{fu2026synthesis}, while RTL-OPT focuses directly on design optimization
and shows tool- and constraint-dependent rankings \cite{lu2026rtlopt}. Security
benchmarks add domain-specific weaknesses and properties
\cite{chen2026hardsecbench}.

These efforts motivate richer evidence, but richer metric lists do not by
themselves define when a comparison is licensed. They evaluate changing
candidates under a selected stack; we hold candidates fixed and perturb the
stack's semantics and gates. A requirement profile declares which evidence is
applicable, how it is produced, which gates precede optimization, how near-ties
and tool failures are handled, and which tool/configuration identities make the
result reconstructable. The experimental object is therefore the mapping from
evidence to an engineering decision. Accordingly, we neither propose a
universal scalar score nor claim a larger RTL corpus.

\section{Tool-Grounded Requirement Profiles}

For design $d$, context $c$, and requirement $r$, define

\begin{equation}
Q_{d,c,r}=\langle D_r(\theta_{d,c}), E_{r,d,c}, C_{r,d,c}\rangle,
\end{equation}

where $D$ states meaning, scope, applicability, and budget; $E$ is an
executable evidence procedure; and $C$ maps two valid evidence records to a
comparison relation. A complete bundle is

\begin{equation}
B_{d,c}=\langle G_{d,c},\{Q_{d,c,r}\},H_{d,c},U_{d,c},\Pi_{d,c}\rangle .
\end{equation}

$G$ is the common functional-validity gate; $H$ defines precedence across
requirements; $U$ defines tolerance, timeout, invalid, and inconclusive
semantics; and $\Pi$ records source, tool, configuration, workload, and seed
identities. The output relation is a partial order---$A\succ B$, $B\succ A$,
equivalent within evidence, trade-off, or inconclusive---rather than an
unqualified scalar rank.

\begin{figure}[t]
\centering
\includegraphics[width=\columnwidth]{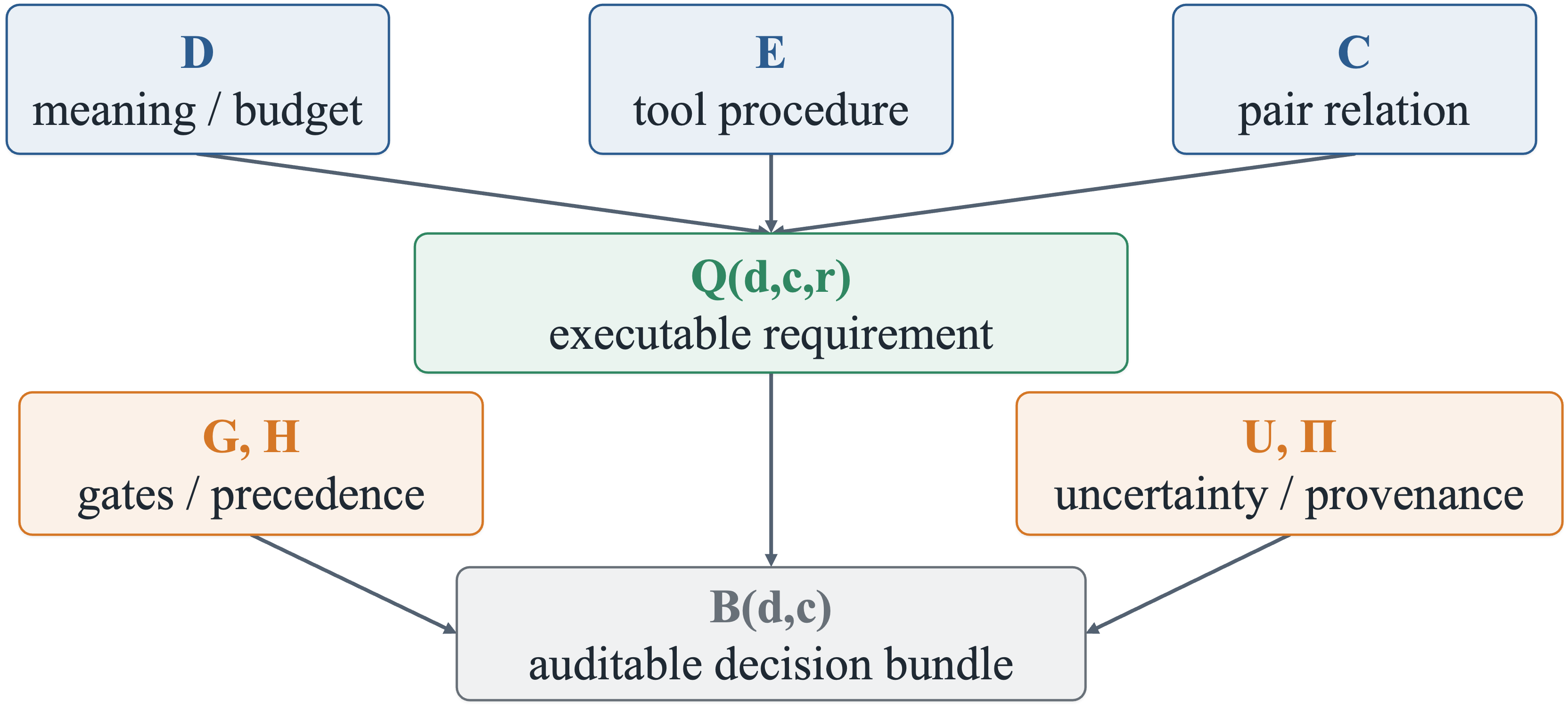}
\caption{A profile binds a requirement to executable evidence and a decision
relation; bundle-level gates, uncertainty, and provenance determine whether a
comparison is licensed.}
\label{fig:schema}
\end{figure}

Figure~\ref{fig:schema} is the general bundle instantiated by the complete C5
evaluator; C0--C4 later remove, replace, or scalarize selected bundle elements.

Three design choices matter. First, evidence records are candidate-independent
and typed: a tool error is not a failed requirement, and missing evidence is
not a tie. Second, feasibility gates precede optimization. A smaller design
that violates the matched behavior is not an improvement. Third, provenance is
part of validity, not supplemental metadata: a metric without its source,
constraints, tool revision, workload, and seed is not accepted as comparable.

\section{Prospective Evaluator Audit}

\subsection{Designs and Controlled Pairs}

The pilot contains six independently authored, paper-only RTL designs and two
controlled variants per design. The security family comprises a secure CSR and
a key vault; the protocol family comprises a ready/valid FIFO and a
round-robin arbiter; the transactional family comprises a request/response
multiplier and a streaming CRC8. Variants inject access-control, lifecycle,
handshake, fairness, response-timing, or transaction-boundary regressions while
remaining syntactically valid and generically synthesizable. Every pair has a
black-box matched oracle independent of candidate internals. The reference is
$A$ and the injected variant is $B$, so the supported matched decision is
$A\succ B$ for all twelve pairs.

This construction measures evaluator sensitivity, not natural defect
prevalence. Using two pairs per design prevents one hand-picked defect from
being the sole unit, while clustering uncertainty by design avoids treating
pairs as independent designs.

\begin{table*}[t]
\centering
\caption{Frozen controlled pairs. $A$ is the reference and $B$ is the injected
variant. Every $B$ passes syntax, elaboration, and generic synthesis; C5
rejects it on the listed distinguishing requirement.}
\label{tab:pairs}
\scriptsize
\begin{tabular}{@{}llll@{}}
\toprule
Pair & Design/family & Injected variant & Distinguishing requirement \\
\midrule
RP001\_P1 & secure CSR / security & write privilege bypass & user write to privileged register is denied \\
RP001\_P2 & secure CSR / security & reversible lock & lock is monotonic until reset \\
RP002\_P1 & key vault / security & debug key leak & debug output contains no key bits \\
RP002\_P2 & key vault / security & load over zeroize & zeroize has priority over load \\
RP003\_P1 & ready/valid FIFO / protocol & simultaneous-count bug & push/pop preserves occupancy \\
RP003\_P2 & ready/valid FIFO / protocol & pop without ready & stalled output remains stable \\
RP004\_P1 & RR arbiter / protocol & fixed priority & bounded round-robin fairness \\
RP004\_P2 & RR arbiter / protocol & advance on stall & grant stable until accepted \\
RP005\_P1 & multiplier / transactional & unsigned multiply & signed multiplication correctness \\
RP005\_P2 & multiplier / transactional & response pulse & response held under backpressure \\
RP006\_P1 & streaming CRC8 / transactional & wrong polynomial & CRC8/ATM reference vectors \\
RP006\_P2 & streaming CRC8 / transactional & response pulse & response held under backpressure \\
\bottomrule
\end{tabular}
\end{table*}

\subsection{Evidence and Decision Reconstruction}

Each condition emits a typed record for both candidates and a pair-decision
record naming the rule and reason code. Syntax and elaboration use Icarus
Verilog; generic cell counts use a pinned Yosys invocation. C5 executes the
case's candidate-independent testbench before consulting synthesis evidence.
The runner never infers missing fields from filenames or logs: invalid,
timeout, and absent evidence remain explicit states.

Evidence production and decision reconstruction are separate. Candidate
evidence is hashed; each condition receives only manifest-authorized fields,
preventing an ablated evaluator from seeing the matched result. The decision
table can be reconstructed from records, while the clean rerun regenerates all
evidence to test end-to-end determinism.

\subsection{Frozen Conditions}

Table~\ref{tab:conditions} lists the prospective conditions. C5 is the proposed
complete operational evaluator: it executes the matched oracle and applies
feasibility before optimization. C0--C4 are controlled baselines or
perturbations derived by removing, replacing, or scalarizing parts of C5. C0
and C1 retain only common eligibility checks. C2 chooses the lower generic
Yosys cell count with zero tolerance. C3 substitutes the wrong generic
reachability profile, and C4 removes the pair's distinguishing gate from the
otherwise matched profile.

\begin{table}[t]
\centering
\caption{Frozen evaluator conditions. C5 is the complete proposed evaluator;
C0--C4 are baselines or controlled perturbations.}
\label{tab:conditions}
\scriptsize
\begin{tabular}{@{}lll@{}}
\toprule
ID & Evidence retained & Decision rule \\
\midrule
C0 & syntax, elaboration & tie if both pass \\
C1 & C0 + generic synthesis & tie if both pass \\
C2 & C1 + cell count & minimize cells \\
C3 & mismatched reachability & wrong context \\
C4 & matched minus key gate & ablated profile \\
C5 & matched oracle + synthesis & feasibility first \\
\bottomrule
\end{tabular}
\end{table}

The protocol, condition manifest, candidates, and analysis plan were frozen at
commit \texttt{20c87ce} before confirmatory execution. A checksum verifier
guards 53 direct and nested frozen artifacts. No post-freeze repair or pair
substitution was allowed.

\begin{figure}[t]
\centering
\includegraphics[width=\columnwidth]{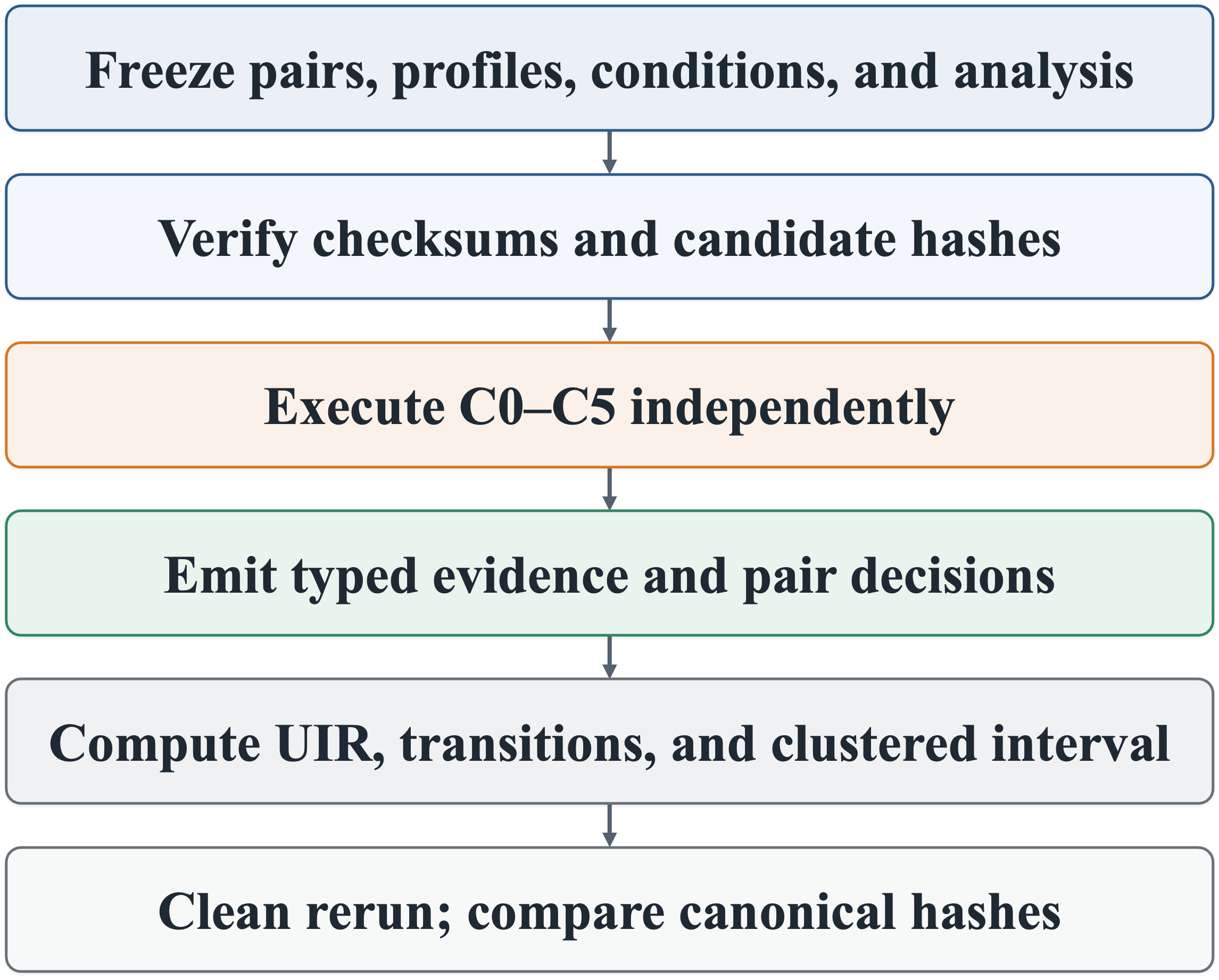}
\caption{Confirmatory execution. Candidate pairs are fixed; only the evaluator
condition changes.}
\label{fig:flow}
\end{figure}

\subsection{Metrics and Analysis}

For pair $i$, let $M_i$ be the matched C5 decision and $S_i$ the scalar C2
decision. An unsupported improvement occurs when C2 declares candidate $B$
better but C5 does not license that claim:

\begin{equation}
\uir_{C2}=\frac{1}{N}\sum_{i=1}^{N}
\mathbf{1}[S_i=(B\succ A)\land M_i\ne(B\succ A)].
\end{equation}

We distinguish a \emph{gate reversal}, where C5 establishes $A\succ B$, from
an \emph{unresolved scalar claim}, where C5 is inconclusive. The latter is not
asserted to be a regression; it remains unsupported because mandatory evidence
does not license the scalar preference. In the controlled study every matched
outcome is $A\succ B$, so its UIR contains only gate reversals.

We also report loss of required distinction, defined as any condition decision
unequal to $M_i$, the full transition counts into C5, matched-regression
detection, profile executability, and decision coverage. Because $N=12$ is
small, we emphasize counts and estimation: a Wilson binomial interval and a
10,000-resample percentile bootstrap clustered over the six designs with seed
20261008 \cite{wilson1927probable}. Family results are always shown with their
denominators.

\section{Confirmatory Results}

All 72 planned decisions executed, and all twelve profiles were executable.
C2 produced unsupported improvement for 6/12 pairs, so
$\uir_{C2}=0.50$. The Wilson 95\% interval is $[0.254,0.746]$ and the
design-cluster bootstrap interval is $[0.25,0.75]$. Each family contributes
2/4 unsupported improvements; the aggregate is therefore not driven by a
single requirement family. C5 detects 12/12 injected regressions.

\begin{table}[t]
\centering
\caption{Confirmatory evaluator audit. LRD is loss of the required C5
distinction; UI is unsupported improvement.}
\label{tab:primary}
\scriptsize
\begin{tabular}{@{}lrrl@{}}
\toprule
Condition & LRD & UI & Interpretation \\
\midrule
C0 syntax & 12/12 & 0/12 & all false ties \\
C1 generic & 12/12 & 0/12 & all false ties \\
C2 scalar & 7/12 & 6/12 & 6 reversals, 1 tie \\
C3 mismatched & 12/12 & 0/12 & all false ties \\
C4 ablated & 12/12 & 0/12 & all false ties \\
C5 matched & 0/12 & 0/12 & detects 12/12 regressions \\
\bottomrule
\end{tabular}
\end{table}

Across all 60 non-matched decisions, the transition matrix into C5 contains
five $A\succ B\rightarrow A\succ B$ decisions, six
$B\succ A\rightarrow A\succ B$ reversals, and 49
tie$\rightarrow A\succ B$ transitions. Thus scalarization is not the only
failure mode: omission frequently erases a required distinction without
explicitly preferring the defective variant.

\begin{table}[t]
\centering
\caption{Pair-level C2 audit. Cell counts are generic Yosys cells for reference
$A$ and injected variant $B$. Bold rows are unsupported improvement reversals;
C5 returns $A\succ B$ for every row.}
\label{tab:pairresults}
\scriptsize
\begin{tabular}{@{}lrrll@{}}
\toprule
Pair & Cells $A$ & Cells $B$ & C2 decision & C2 $\rightarrow$ C5 \\
\midrule
RP001\_P1 & 352 & 353 & $A\succ B$ & retained \\
\textbf{RP001\_P2} & \textbf{352} & \textbf{351} & $\mathbf{B\succ A}$ & \textbf{reversal} \\
RP002\_P1 & 391 & 423 & $A\succ B$ & retained \\
\textbf{RP002\_P2} & \textbf{391} & \textbf{390} & $\mathbf{B\succ A}$ & \textbf{reversal} \\
RP003\_P1 & 831 & 831 & equivalent & tie $\rightarrow A\succ B$ \\
RP003\_P2 & 831 & 832 & $A\succ B$ & retained \\
\textbf{RP004\_P1} & \textbf{82} & \textbf{17} & $\mathbf{B\succ A}$ & \textbf{reversal} \\
\textbf{RP004\_P2} & \textbf{82} & \textbf{80} & $\mathbf{B\succ A}$ & \textbf{reversal} \\
\textbf{RP005\_P1} & \textbf{1822} & \textbf{1547} & $\mathbf{B\succ A}$ & \textbf{reversal} \\
RP005\_P2 & 1822 & 1824 & $A\succ B$ & retained \\
RP006\_P1 & 84 & 92 & $A\succ B$ & retained \\
\textbf{RP006\_P2} & \textbf{84} & \textbf{83} & $\mathbf{B\succ A}$ & \textbf{reversal} \\
\bottomrule
\end{tabular}
\end{table}

Table~\ref{tab:pairresults} exposes why aggregate score differences are
insufficient. Four defective variants save only one or two cells; RP004\_P1
and RP005\_P1 show large apparent savings by eliminating required behavior. A
tolerance could suppress small reversals but neither restore lost behavior nor
make the metric requirement-aware. Five C2 decisions agree with C5 only because
the reference also has fewer cells; only the matched oracle licenses the result.

\begin{figure}[t]
\centering
\begin{tikzpicture}[x=.43cm,y=.34cm,font=\scriptsize]
  \draw[->] (0,0) -- (13,0) node[right] {pairs};
  \foreach \y/\lab in {1/C5,2/C4,3/C3,4/C2,5/C1,6/C0}{
    \node[anchor=east] at (-.2,\y) {\lab};
  }
  \fill[profilegreen!80] (0,0.65) rectangle (12,1.35);
  \fill[profileorange!80] (0,1.65) rectangle (12,2.35);
  \fill[profileorange!80] (0,2.65) rectangle (12,3.35);
  \fill[profileorange!80] (0,3.65) rectangle (7,4.35);
  \fill[profilegray!45] (7,3.65) rectangle (12,4.35);
  \fill[profileorange!80] (0,4.65) rectangle (12,5.35);
  \fill[profileorange!80] (0,5.65) rectangle (12,6.35);
  \foreach \x in {0,2,4,6,8,10,12}{\draw (\x,-.12)--(\x,.12) node[below=2pt]{\x};}
  \node[anchor=west] at (0,7.1) {\color{profileorange}\rule{2.2mm}{2.2mm} LRD \quad
    \color{profilegray}\rule{2.2mm}{2.2mm} retained \quad
    \color{profilegreen}\rule{2.2mm}{2.2mm} matched};
\end{tikzpicture}
\caption{Pairs losing the required matched distinction. C2 retains the
distinction for five pairs but reverses six others; the other incomplete
conditions collapse all twelve comparisons.}
\label{fig:decisions}
\end{figure}
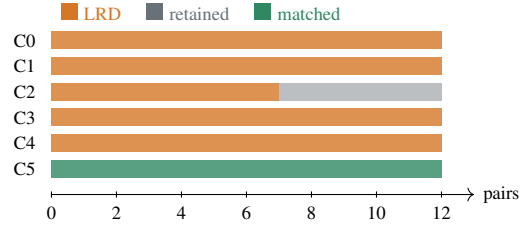

The clean rerun exactly reproduced the canonical decision rows
(SHA-256 prefix \texttt{61f62263}) and analysis (prefix
\texttt{15dd86a6}). These hashes exclude only ephemeral paths and timestamps,
not evidence or decisions.

\subsection{Robustness and Interpretation}

Security, protocol, and transactional designs each contribute 2/4 to the 6/12
result. The wide clustered interval reflects sensitivity to removing a design,
so we make no null-hypothesis significance claim. Clustering is planned because
two variants share design, oracle, and tool context. UIR counts false preference
for $B$; loss-of-distinction also counts RP003\_P1's false tie. C0, C1, C3, and
C4 create no reversal here but erase 12/12 distinctions, so both metrics are
needed.

Because zero tolerance could make C2 appear artificially brittle, we performed
a post-freeze exploratory sensitivity analysis using relative cell-count
tolerances. Unsupported improvement remains 4/12 at 0.5--1\%, 3/12 at 2\%, and
2/12 at 5--10\%; the two remaining reversals remove 79.3\% and 15.1\% of the
reference cell count by deleting required behavior. Tolerance also converts
some supported scalar preferences into ties: loss of required distinction is
10/12 at 0.5--5\% and 12/12 at 10\%. Thus a near-tie policy can suppress small
reversals but cannot reconstruct an omitted requirement gate. At the frozen
zero tolerance, leave-one-design-out UIR ranges from 4/10 to 6/10, so no single
design creates the 6/12 result. These analyses are robustness checks, not part
of the preregistered confirmatory estimand.

\subsection{Text-Only External Validation}

After completing the controlled study, we froze a separate external-validation
protocol before inspecting candidate-level outcomes. It includes every public
text-only candidate in the final RTLLM v1.1 snapshot: 29 tasks, two upstream
model labels, and five trials per label, for 290 candidate/reference pairs
\cite{lu2024rtllm}. This extension neither modifies the 72-decision
confirmatory matrix nor introduces a multimodal condition. Exact provider
snapshots and generation parameters are not independently reconstructable, so
the inherited labels are blocking variables rather than objects of comparison.

For a deliberately conservative contrast, C2 minimizes generic Yosys cells
without functional evidence. C5 applies the upstream task testbench first and
uses the same cell-count relation only when both implementations pass mandatory
functional and synthesis gates. Thus a reversal isolates the missing gate, not
a different optimization objective. Four tasks had invalid reference evidence
in the portable Icarus/Yosys pipeline and were retained in the artifact but
excluded from the estimand: two reference/testbench identifier collisions
(\texttt{div\_16bit}, \texttt{pe}), one upstream-oracle failure
(\texttt{radix2\_div}), and one portable-synthesis failure
(\texttt{synchronizer}). The remaining 25 tasks yielded 215 pairs with
reference validity and cell counts for both sides.

\begin{table}[t]
\centering
\caption{Evidence coverage and C2-to-C5 transitions in the public text-only
cohort. The 215 evaluable pairs are clustered within 25 tasks.}
\label{tab:external}
\scriptsize
\begin{tabular}{@{}llrl@{}}
\toprule
\multicolumn{4}{@{}l}{\textit{Evidence coverage}} \\
\multicolumn{2}{@{}l}{Record/stage} & Count & Disposition \\
\midrule
\multicolumn{2}{@{}l}{All source candidates} & 290 & retained \\
\multicolumn{2}{@{}l}{Syntax/elaboration pass} & 207 & candidate evidence \\
\multicolumn{2}{@{}l}{Functional pass} & 96 & candidate evidence \\
\multicolumn{2}{@{}l}{Generic synthesis pass} & 252 & candidate evidence \\
\multicolumn{2}{@{}l}{Primary-evaluable pairs} & 215 & C2/C5 contrast \\
\midrule
\multicolumn{4}{@{}l}{\textit{Decision transitions}} \\
C2 & C5 & Pairs & Interpretation \\
\midrule
$A\succ B$ & $A\succ B$ & 61 & retained \\
$B\succ A$ & $A\succ B$ & 72 & gate reversal \\
$B\succ A$ & $B\succ A$ & 17 & licensed improvement \\
$B\succ A$ & inconclusive & 9 & unsupported unresolved \\
equivalent & $A\succ B$ & 14 & lost distinction \\
equivalent & equivalent & 42 & retained tie \\
\bottomrule
\end{tabular}
\end{table}

Primary evaluability requires a valid reference and scalar synthesis evidence
for both sides; it does not require $B$ to pass function. A recorded functional
failure is matched gate evidence that licenses C5 to reject $B$, whereas a
timeout or tool error remains unresolved. Thus the 215-pair denominator is
consistent with 96 functional passes rather than restricted to them.

C2 prefers the generated candidate in 98 pairs, but C5 licenses only 17 of
those preferences. The other 81 are unsupported improvements, giving
$\uir=81/215=37.7\%$ (descriptive Wilson 95\% interval 31.5--44.3\%;
10,000-resample task-cluster interval 24.8--51.4\%). Of these, 72 are explicit
functional-gate reversals and nine remain inconclusive because mandatory
functional evidence is unresolved. A clean rerun reproduces all 290 canonical
decision rows exactly (SHA-256 prefix \texttt{975bc50f}). The result is not a
model leaderboard or prevalence estimate; it shows that the mechanism observed
with controlled regressions also occurs in a complete public text-only cohort.
Unsupported improvements occur in 18/25 reference-valid tasks, and no task
contributes more than 9/81 events. This breadth reduces the risk that the
aggregate is an artifact of one task, while task-level rates from 0--100\%
reinforce why the clustered interval, rather than an i.i.d. interval, is the
preferred uncertainty summary.

\section{Physical-Design Feasibility Studies}

The primary experiment audits the evaluator and does not require place and
route. We nevertheless test whether profiles can bind physical context and
reject invalid energy evidence. Two transactional designs were implemented in
a pinned AArch64 OpenROAD-flow-scripts environment using Nangate45
\cite{ajayi2019openroad}. These 26 runs and 2,938 registered artifacts are
\emph{non-confirmatory feasibility evidence}; they were collected before the
protocol freeze and are not used in the primary UIR estimate.

RP005 compares a one-cycle multiplier with a 16-cycle iterative
implementation. RP006 compares a one-cycle parallel CRC8 with a 32-cycle
serial implementation. Table~\ref{tab:ppa} shows that the area relation varies
with clock context and that feasibility changes at the aggressive boundary.
All retained runs report zero detailed-route DRC errors. LVS was not run
because no compatible deck was available, so these results are not signoff.

\begin{table}[t]
\centering
\caption{Non-confirmatory ORFS feasibility contexts. Ratio is iterative/single
for RP005 and serial/parallel for RP006.}
\label{tab:ppa}
\scriptsize
\begin{tabular}{@{}llrrl@{}}
\toprule
Case & Context & Period (ns) & Area ratio & Feasible \\
\midrule
\multirow{3}{*}{RP005} & relaxed & 1.6875 & 0.850 & both \\
 & nominal & 1.3500 & 0.783 & both \\
 & aggressive & 1.2150 & 0.670 & iterative only \\
\midrule
\multirow{3}{*}{RP006} & relaxed & 0.5500 & 2.119 & both \\
 & nominal & 0.4750 & 2.019 & both \\
 & aggressive & 0.4500 & 2.002 & serial only \\
\bottomrule
\end{tabular}
\end{table}

Five-seed nominal repeatability gives area and delay range-over-mean of 0.278\%
and 0.602\% for the RP005 one-cycle design. For RP006 parallel CRC8, the
corresponding values are 1.098\% and 0.990\%; qualified active energy varies by
1.767\% with a mean of 3.933 pJ/operation. These observations motivate
metric-specific resolution floors rather than assuming exact physical
determinism.

Most importantly, RP006 demonstrates invalid evidence handling. The parallel
trace passes boundary coverage and yields an active/idle dynamic-power ratio of
35.378. The serial trace yields 0.845, failing the prerequired discrimination
check; its computed energy is therefore excluded from ranking. The profile
returns an inconclusive energy relation rather than converting a plausible
number into a claim. RP005 energy values are likewise retained only as
diagnostics because their activity-qualification rule was not frozen.

\section{Discussion}

\textbf{What the pilot establishes.} The same frozen RTL pair can transition
from a false tie or apparent improvement to a matched regression solely because
the evaluation contract changes. The effect crosses security, protocol, and
transactional families in this controlled set. Requirement profiles make that
contract explicit enough to preregister, execute, and reconstruct. The external
cohort observes the same missing-gate mechanism in 18/25 reference-valid tasks,
showing that it is not confined to the injected variants.

\textbf{Implications for RTL leaderboards.} A benchmark should publish not
only tests and aggregate scores but also applicability, feasibility precedence,
decision relations, uncertainty floors, invalid/inconclusive policies, and
tool provenance. If a single scalar is reported, its license must follow from
the profile rather than from convenience. Candidate generation and evaluator
configuration should be versioned as separate experimental factors.

\textbf{Adoption protocol.} A benchmark can use the method without adopting
our cases: enumerate applicable requirements; bind each to typed executable
evidence; predefine feasibility, pair relations, tolerances, and unresolved
outcomes; then freeze candidates and audit plausible profile perturbations.
Unstable conclusions should be labeled profile-dependent, not averaged away.

\textbf{Why not one weighted score?} Weights cannot repair an absent gate.
Trading a required violation against area changes the decision semantics.
Partial orders instead optimize within the feasible set and preserve
incomparable trade-offs; scalar summaries are permitted only after the profile
establishes comparability and a value model.

\textbf{Threats to construct validity.} Controlled variants may make the
obligation clearer than real generated failures, and stronger learned or
multi-metric judges may fail differently. Prospective regressions and oracles,
separate evidence/decision stages, and pair-level transitions limit circularity.

\textbf{Threats to internal validity.} A shared implementation bug could affect
both candidates. Independent black-box oracles, frozen hashes, isolated
outputs, logs, and a clean rerun reduce this risk. Testbenches establish the
frozen obligations, not all possible behavior.

\textbf{Threats to external validity.} The controlled study still contains six
small designs. The public extension broadens evidence to 290 historical
text-only outputs over 29 RTLLM tasks, but exact provider snapshots and
generation parameters are unavailable, four reference tasks are invalid in the
portable pipeline, and the supplied testbenches are not exhaustive. Its UIR is
therefore audit sensitivity, not current ecosystem defect prevalence.

\textbf{Physical-evidence limits.} The PPA study uses one open 45-nm platform,
has no LVS result, and is non-confirmatory. Boundary and propagation diagnostics
do not eliminate low direct VCD annotation. Five seeds estimate local flow
variability, not process or cross-tool variation.

\textbf{Publication boundary.} This study contains no multimodal prompt
condition, diagram input, representation comparison, or generator leaderboard.
A future journal study may reuse the evaluator methodology while asking the
separate question of how specification representations affect generated RTL.
No candidate or result from that future experiment is included here.

\textbf{Responsible use.} The authors remain responsible for designs, oracles,
statistics, claims, citations, and release decisions, including any automated
assistance. The artifact does not establish deployment safety.

\section{Reproducibility}

The artifact records frozen manifests and checksums, candidate identities and
source hashes, condition-specific evidence, all 72 decisions, analysis seed and resampling
count, tool versions, and canonical rerun hashes. The external extension pins
the upstream RTLLM commit, hashes all 290 candidates and task assets, preserves
invalid/unresolved outcomes, and reproduces its canonical decision hash on a
clean rerun. The evaluator runner verifies each freeze before reconstructing
decisions, and the planned task-level summary is derived from the canonical
rows. Compact PPA bundles include commands, source/configuration hashes,
clock contexts, timing, area, DRC, activity, power, and seed summaries.
\ifarxivversion\else
\fi

\section{Conclusion}

RTL evaluation is itself an experimental treatment. Tool-grounded requirement
profiles make its assumptions executable and auditable, while controlled
profile perturbations reveal when a conclusion depends on missing requirements
or mismatched context. In this frozen pilot, a generic scalar evaluator makes
six unsupported improvement claims in twelve pairs, while the matched profiles
detect every injected regression. A separate complete public text-only cohort
finds 81 unsupported scalar improvements among 215 evaluable pairs, including
72 functional-gate reversals. Physical-design studies further show that context
and evidence qualification change what can be concluded. The practical lesson
is direct: freeze and audit the evaluator before trusting the ranking.

\ifdacreferencespage\clearpage\fi
\bibliographystyle{IEEEtran}
\bibliography{references}

\end{document}